\documentclass[aps, prd, superscriptaddress, nofootinbib]{revtex4}
\usepackage{graphicx}
\usepackage{dcolumn}
\usepackage{amssymb}
\usepackage{amsmath,amssymb,amsfonts}

\begin{document}

\newcommand{\be}{\begin{equation}}
\newcommand{\ee}{\end{equation}}
\newcommand{\simless}{\lower.5ex\hbox{$\; \buildrel < \over \sim\;$}}
\newcommand{\simgreat}{\lower.5ex\hbox{$\; \buildrel > \over \sim\;$}} 
\renewcommand{\thefootnote}{\fnsymbol{footnote}}
\newcommand{\econ}{ {\rm e} } 
\newcommand{\mplan}{ M_{\rm planet}} 
\newcommand{\mplanck}{ M_{\rm pl}} 
\newcommand{\mstar}{ M_\star }
\newcommand{\mequal}{ M_{\rm eq} }
\newcommand{\mgal}{ M_{\rm gal} }
\newcommand{\tempeq}{ T_{\rm eq} }
\newcommand{\timeeq}{ t_{\rm eq} }
\newcommand{\rhoeq}{\rho_{\rm eq}} 
\newcommand{\mpro}{ m_{\rm p} }
\newcommand{\melec}{ m_{\rm e} }
\newcommand{\fgal}{f_{\rm gal}} 
\newcommand{\tvir}{t_{\rm vir}} 
\newcommand{\fvir}{f_{\rm vir}} 
\newcommand{\rhovac}{\rho_{\scriptscriptstyle\Lambda} }

\newcommand{\ba}{\begin{eqnarray}}
\newcommand{\ea}{\end{eqnarray}}
\def\bs{\begin{subequations}}
\def\es{\end{subequations}}
\def\a{\alpha}
\def\b{\beta}
\def\g{\gamma}
\def\e{\epsilon}
\def\t{\theta}
\def\s{\sigma}
\def\cE{{\cal E}}
\def\cH{{\cal H}}
\def\cN{{\cal N}}
\def\cV{{\cal V}}
\def\cG{{\cal G}}
\def\p{\partial}
\newcommand{\Eq}[1]{(\ref{#1})}
\def\lp{\ell_{\rm Pl}}

\title{On a Relation of Vacuum Energy to the Hierarchy of Forces}

\author{Stephon Alexander}
\affiliation{Department of Physics, Brown University, Providence, RI 02906, USA}
\author{Laura Mersini-Houghton}
\affiliation{Department of Physics and Astronomy, UNC-Chapel Hill, NC 27599, USA}
\date{\today}

\begin{abstract}
We investigate the relation between vacuum energy and the hierarchy of forces in habitable universes. The hierarchy of forces studied in \cite{fred} was bound by $\frac{\alpha_G}{\alpha} \le 10^{-34}$, using structure formation arguments which  involve the fine structure constant $\alpha$ and the gravitational constant $\alpha_G$. Previously we showed that the requirement of vacuum domination epoch occuring after matter radiation equality time, places a bound on vacuum energy in terms of the density perturbation parameter Q. Here we impose a further condition: we require that at equality time the size of the gas cloud which forms a galaxy at virialization time, be smaller than the horizon size. The latter condition leads to an intriguing  relation whereby vacuum energy is bound by a power law function of the hierarchy of forces. The constraint introduced by the hierarchy of forces on the amount of dark energy is suggestive of an unknown microphysical mechanism that relates vacuum energy to the other constants of nature, specifically to the gravitational and fine structure constants and to their hierarchy.
\end{abstract}


\maketitle

Despite the successes of the standard model of cosmology and of particle physics, there remain some deep mysteries concerning fine tuning of the constant of nature, the selection of the initial conditions of the early universe, the smallness of vacuum energy, the electroweak vacuum instability, and the hierarchy of forces. The difficulty with these fundamental questions, has led a number of investigators to consider anthropic reasoning. Given the major advances in precision cosmology, strides of progress have been made in the last decade in probing the origin of cosmic inflation, a period of vacuum energy domination of the very early universe, which may well lead towards an expanded framework of cosmology, a multiverse. 

On the other hand, theoretical progress with the understanding of vacuum energy and the fine tuning of constants of nature lags behind. We know observationally that our universe has a large hierarchy of forces and is currently dominated by a tiny vacuum energy density.  We do not yet understand why vacuum energy is so small, if it is a constant, or if it has any connection to other constants of nature. Structure formation reasoning at least sheds some light on why the constants of nature are tuned within a range such as to give a large hierarchy of forces. Weinberg \cite{weinberg} argued decades ago that large scale structures can not form if the vacuum energy was greater than the value that we observe today.  This happens because the repulsive nature of vacuum energy counters the gravitational clustering of matter during the epoch of structure formation.  This prediction was made before the observation of an accelerating universe \cite{SN1a}, and seemed to support the anthropic principle, although the latter was challenged in \cite{fredlaura} applying the cosmic heat death argument that DeSitter spaces have in common.

In this work we consider the implications that variations of the other constants of nature would have on vacuum energy by exploring possible connections among them. We find a functional dependence of vacuum energy on the hierarchy of forces when the Q parameter for structure formation, the cooling time and gravitational collapse time of structures, and the bounds on the size of galaxies given by the horizon size at equality, are simultaneously taken into consideration, and the fine structure and gravitational constants of nature $\alpha$, and $\alpha_G$, are allowed to vary.  Specifically, we end up with an intriguing interconnection whereby the cosmological constant is directly limited by the relative values of the fine structure and gravitational constants in any habitable universe.

\it{Warm Up: Structure Formation Considerations}\rm 

We will be making use of the following definitions and astrophysical relations \cite{tegrees, tegmark, fred, whiteost, silk}: the typical star mass $M_{0}=m_{p}\alpha_{g}^{-3/2}$, where $m_{p}$ is the proton mass, and $\alpha_{g}= \frac{G m_{p}^2}{{\bar h}c}$ is the dimensionless gravitational constant.A typical planet has a mass $M_{0p} =m_{p}(\alpha/\alpha_{g})^{3/2}$ with $\alpha$ the fine structure constant. The constant $a_R =\frac{\pi^2}{6}$ enters the expression for the radiation energy density below.

 The Friedman equation for the expansion of the universe is
\be
\left( {{\dot a} \over a} \right)^2 = \Omega_{\lambda} + 
\Omega_M a^{-3} + \Omega_R a^{-4} \,. 
\ee
where $\Omega_{\lambda}, \Omega_{M}, \Omega_R$ are the vacuum, matter and radiation energy density components respectively relative to the critical energy density of the universe.
We also have the constraint 
\be
\Omega_{\lambda} + \Omega_M + \Omega_R = 1 \,, 
\ee
where we have assumed that the universe is spatially flat. The time of matter radiation equality is denoted by $t_{eq}$,  and $\Omega_b$ is the baryonic matter energy density..  Perturbation growth starts after matter domination at $t_m > t_{eq}$ and it stops at vacuum domination epoch denoted by $t_{\lambda}$. The universe has a total energy density denoted by $\rhoeq$ at the epoch of equality.

The temperature of the universe at the 
epoch of equality is given by the condition that the radiation energy per photon is equal to the matter energy per proton \cite{tegrees,tegmark}, which leads to the expression
\be
k \tempeq = \eta (\mpro c^2) {\Omega_M \over \Omega_b}\left(1 + \frac{\Omega_{\lambda}}{\Omega_{m}}\right) \,,
\label{tequal} 
\ee
where the parameter $\eta$ is the baryon to photon ratio, and the energy density of radiation goes as $T_{eq}^{4}$. For our universe $\eta = 10^{-9}$. Another parameter often used is the ratio of matter energy to photon density, related to the temperature at equality by $\xi = T_{eq}$.

If vacuum energy density is larger than that of matter, then no structures form in the universe, since vacuum energy will dominate before structures have a chance to grow and condense.
So, it is reasonable to assume that vacuum energy dominates after structure forms $\rho_{\lambda} \ll \rho_{m}$, which allows us to ignore the correction term $\frac{\Omega_{\lambda}}{\Omega_m} \ll 1$ in \ref{tequal}. The energy density at matter radiation equality time $t_{eq}$ is 
\be
\rho_{m} = \rho_{rad} = 2 a_{r} T_{eq}^4
\label{equalityenergy}
\ee

with $a_r$ defined above. The mass $M_{eq}$ enclosed by the Hubble radius at equality time 
\be
t_{eq} = H{z_{eq}}^{-1}
\label{equalitytime}
\ee

 can be estimated from
 
\be
M_{eq} = \rho_{eq} t_{eq}^{3} \approx T_{eq}^{4} t_{eq}^{3}\approx \xi^{2} \alpha_{G} M_{0} (\frac{3}{8 \pi G})^{1/2}
\label{equalitymass}
\ee
which can be further expressed in terms of fundamental constants of nature and stars mass as

\be
\mequal = \left( {5 \over \pi} \right)^{1/2} 
{3 \over 64\pi} \alpha_G^{-3/2} \mpro 
\left( {\mpro c^2 \over k \tempeq} \right)^2
\approx {M_{0} \over 64 \eta^2} 
\left( {\Omega_b \over \Omega_M} \right)^2\,,
\label{equalitymass}
\ee
The Hubble parameter at equality time is obtained from the Friedmann equation 
\be
H(z_{eq}) ^{2} = H_{eq}^{2} = \frac{8\pi G}{3} T_{eq}^{4} (1 + \frac{\rho_{\lambda}}{\rho_m})
\label{equalityhubble}
\ee

Gas virializes and the first structure form at a time $t_{vir} > \gg t_{eq}$. The virialization time is obtained from
\be
t_{vir} \simeq t_{eq} Q^{-3/2} f_{vir}
\label{virialtime}
\ee

where $f_{vir}\simeq 1-1000$ is a dimensionless factor which takes into account variations on structures size that form from the gas cloud at $z_{eq}$. We will consider it to be of order one in what follows, without loss of generality. The energy density at $z_{vir}$ is thus given by
\be
\rho_{vir} = \rho_{eq} (t_{vir}/t_{eq})^2 = \rho_{eq} f_{vir}^{2} Q^{-3}
\label{virialenergy}
\ee

Authors of \cite{fred} showed, using structure formation arguments, that there always exists a hierarchy between the electromagnetic and gravitational forces in habitable universes. This hierarchy is given by a bound that two constants of nature, the fine structure constant and the gravitational constants respectively, must obey
\be
\frac{\alpha_{G}}{\alpha} \ll 10^{-34}
\label{hierarchy}
\ee

if structure were to form. Allowing for vacuum energy in the universe does not change this bound on the hierarchy of forces since for structure to form, vacuum domination time must occur after matter domination to allow perturbation to grow. For this reason, as we mentioned the correction factor $\frac{\rho_{\lambda}}{\rho_m} \ll 1$ which modifies $t_{eq}, \xi, T_{eq}$, that enter in the hierarchy between the gravitational and electromagnetic forces given by Eq. \ref{hierarchy} above, introduces an insignificant correction to this expression that can be safely ignored. Below we continue to use the same bound for the hierarchy of forces given by Eq. \ref{hierarchy}, even in the presence of dark energy.  

Vacuum energy is important though where the perturbation growth is concerned. Perturbations grow during the matter domination  epoch $t_{m} > t_{eq}$, then the first structures form at $t_{vir}$ where $(1 + z_{vir}) \simeq Q$. 

Finally perturbation growth stops at a later time $t_{\lambda} > t_{vir}$ during vacuum energy domination epoch $z_{\lambda}$ given by $(1 + z_{\lambda}) = (\frac{\rho_{\lambda}}{\rho_m})^{1/3}$ . The expression for  $z_{vir}$ comes from the fact that $\delta \simeq a(t)$ and the expression for $z_{\lambda}$ is obtained from the fact that vacuum energy dominates from the time when $\rho_{\lambda}(z_{\lambda}) = \rho_m$, using $\rho_{m} \simeq (1 + z)^3$. As far as perturbations are concerned, their dependence on vacuum energy gives a one parameter family of models for structure formation as a function of vacuum energy, discussed in \cite{fsel}. 


In \cite{fredstephonlaura} we derived a bound that dark energy must satisfy in habitable universes, which comes from the requirement that dark energy must dominates {\it after} the first structures have formed, i.e. after virialization time $t_{vir}$

\begin{equation}
t_{\lambda} \gg t_{vir}
\label{timecondition1}
\end{equation}

Through the relation of $z_{\lambda}, z_{vir}$ to $\rho_{\lambda}, Q$ shown above, namely $(1 + z_{\lambda}) = (\frac{\rho_{\lambda}}{\rho_m})^{1/3}$ and $(1 + z_{vir}) \simeq Q$, requiring $t_{\lambda} \gg t_{vir}$ we have
 
\be
\rho_{eq} Q^3 > \rho_{\lambda} \,. 
\label{conzero} 
\ee
Besides the implication of this bound on dark energy discussed in \cite{fredstephonlaura}, which we will refer to as {\it condition 1}, Eq. \ref{conzero} also quantifies our argument that indeed $\frac{\rho_{\lambda}}{\rho_{eq}} \ll 1$ for any values of $(\alpha, \alpha_{G},Q)$, therefore corrections of this order in the expressions Eqn.\ref{tequal, equalityhubble,equalitymass} can be ignored.

\it{Relation between Vacuum Energy and the Hierarchy of Forces}\rm

Besides $Q$, we allow the fundamental constants $\alpha, \alpha_{G}$ to float in their allowed range found in \cite{fred} which spans $6-20$ order of magnitude, for as long as they obey the hierarchy condition given by Eqn.\ref{hierarchy}. In our universe the Q parameter is measured to be $Q \sim 10^{-5}$ \cite{cobe,wmap,planck}. On general grounds $Q\ll1$ in habitable universes in order for the universe not to overclose and for it to be around long enough for structures to form \cite{tegrees,coppess}, and for cooling to occur. 


Here we discuss a different constraint from {\it condition 1} of our previous work \cite{fredstephonlaura} given by Eq. \ref{conzero}. The new constraint is placed at the epoch of matter radiation equality instead of dark energy domination $t_{\lambda}$. As we see below both constraints stem from two simple conditions on timescales involved in the evolution of habitable universes.

We require that virial time happens after equality time and the largest scale that can make a galaxy at $t_{vir}$ from the gas cloud of the $t_{eq}$ time be smaller than the horizon size of the universe. Let us refer to it as {\it condition 2}. In this manner, from matter domination era fluctuations with comoving wavelengths $d \simeq a/(k_d)$ that cross the horizon at time $a_d$ and have wavenumbers $k_{d} = H a_{d}$  can start to grow from $z_{eq}$ and up to the time when they virialize, then condense and become nonlinear around $z_{vir}$. Our new requirement is a simple statement on ordering of time scales by requiring that virialization must occur after matter radiation equality epoch, and on sizes of potential galaxies that condense later, originating from the gas cloud of equality time

\be
t_{vir} \ge t_{eq}
\label{timecondition2}
\ee

or equivalently $z_{vir} < z_{eq}$, where again we use $Q \simeq (1+ z_{vir})$. Now the wavelength of fluctuation that crosses the horizon of size $H(eq)^{-1} \simeq t_{eq}$ at time $z_{eq}$ is $d_{max} \simeq t_{eq}/(1 + z_{eq})$. \footnote{We refrain from using the symbol $\lambda$ for the wavelengths since $\lambda$ is often used to denote vacuum energy}.  The wavelength fluctuation that crosses the horizon at $z_{vir}$ is $R_{max} = t_{eq} / (1 + z_{vir})$. Of course there is a whole bunch of fluctuations that cross the horizon at any time $z_d$ between equality time and virial time $ z_{vir} < z_{d} < z_{eq}$ with wavelengths that scale as $d \simeq t_{eq}/ (1 + z_{d})$. All of these fluctuations are inside our horizon by virial time, growing and condensing, virializing and making structures. The size of any fluctuation of wavelength $d$ as it crosses the horizon at some time $z_d$, with the potential to condense and make a galaxy at a later time $z_{vir}$ of size $R_g$,  will be smaller than the size of the fluctuation that enters at $z_{vir}$ and condenses to make a galaxy of size $R_{max}$ since using the scaling with time of these sizes, and our condition 2 given by Eq. \ref{condition2}, we have
\be
R_{g}/ R_{max} = (1 + z_{vir}) / (1 + z_{d}) < 1
\label{sizes}
\ee

because $z_{vir} < z_d$. The lowest value for $z_d$ is $z_{eq}$ since $z_{eq} \le z_{d} <  z_{vir}$, thus in the above equation Eq.\ref{sizes} replacing $z_d$ with $z_{eq}$ makes the bound on $Q$ even stronger.
   
The last Eq.\ref{sizes} tells us that fluctuations crossing before virial time which may condense later  will have a size $\frac{t_{eq}}{(1+ z_d)} = R_{g} \le R_{max} = \frac{t_{eq}}{(1+z_{vir})}$. Let us initially view our second condition as a bound on $Q$. (Further below we invert it in favor of a vacuum energy relation).

\be
(1+ z_{vir}) \simeq Q \simeq \frac{t_{eq}}{R_{max}} < \frac{t_{eq}}{R_{g}}
\label{condition2Q}
\ee
 
The case of wavelengths with size $R_{g} = R_{max}$ is the limiting case since $Q \simeq 1/R_{max} \simeq (1+z_{vir})$. 
As we can see from Eq. \ref{condition2Q} our {\it condition 2} of Eq.\ref{timecondition2} leads to an upper bound on $Q$. Meanwhile {\it condition 1} of Eqs. \ref{timecondition1, conzero} provides a lower bound on $Q$. 
Putting together both requirements Eqs. \ref{timecondition2, timecondition1}, results in an upper and lower bound on variations of perturbations $Q$, which is a simple consequence of ordering of time scales in universes where structures can form. Namely, from

\be 
t_{\lambda} > t_{vir} > t_{eq}
\ee

which is re expressed as the following bounds on $Q$

\be
(\frac{\rho_{\lambda}}{\rho_{m}})^{1/3} < Q (\simeq t_{eq}/ R_{max}) < \frac{t_{eq}}{R_{g}}
\label{finalbounds}
\ee

Below we make use of the expression for $R_{galax}$ rewritten in terms of fundamental constants \cite{silk} in order to relate vacuum energy to those constants. 

\footnote{A back of the envelope calculation if using $R_{gal} = R_{max}$ shows that  $\rho_{\lambda} \simeq \alpha_{g} ^{9/2}$ corresponds to the largest fluctutation entering at virial time which may condense into a galaxy of size close to $R_{max}$. We know $R_g \le R_{max}$. The expression used below for $R_{gal}$ would correspond to the largest galaxy one can make in the universe. In reality most galaxies will originate from fluctuations with size $R_{gal}$ that are somewhere in the range given by Brehstrahhlung and Compton scattering bounds, and $R_{max}$.}


Below we are interested in rewritting the upper bound on $Q$ of Eq. \ref{finalbounds} in favor of vacuum energy. We use the expressions for $Q$ and $R_{gal}$ in terms of constants of nature. This will allow us to explore the impact that variations of the fine structure constant $\alpha$ and of the gravitational constant $\alpha_G$, which are bound by an hierarchy of forces, have on the vacuum energy density $\rho_{\lambda}$
Instead of using the observed
values for $(Q,G,\eta)$ in our universe, we let these constants of nature to float within their allowed bounds discussed in \cite{fred}. From Eq.\ref{finalbounds} rewritten as an inequality $\rho_{\lambda}$ need to satisfy, we find that the vacuum energy is bound by a function of these constants of nature and their hierarchy. The particular constants of nature varied here, the fine structure constant $\alpha$ and the gravitational constant $\alpha_G$ are coupling constants for the electromagnetic and gravitational interactions respectively. Therefore a hierarchy in their values given by Eqn.\ref{hierarchy} is in fact an hierarchy of their respective forces.

Let us now consider the effect of vacuum energy on cooling and structure formation and show how the condition placed at $t_{eq}$  Eq. \ref{timecondition2,} leads to a relation between the energy of the vacuum and the hierarchy of gravitational to electromagnetic forces. The mass scale contained within the cosmological horizon at the time
of equality $M_{eq}$ of Eq. \ref{equalitymass} plays an important role in considerations of structure
formation. This mass scale can be written in the form where the temperature at equality is given by equation (\ref{tequal}). Meanwhile the typical mass and radius scale for galaxy formation is not well understood \cite{silk,carr,whiteost,reesost}, as nonlinear physical processes occur during galaxy formation. Including the condition that the gas can cool within a free-fall time, gives an approximate expression $\mgal = \fgal^2 \alpha_G^{-2} \alpha^5 
( {\mpro \over \melec})^{1/2} \mpro $. The dimensionless factor $\fgal$ takes into 
account the fact that not all galaxies will have the same mass. Galaxies span a range 
of masses (at least in our universe) around $\mgal$ that extend a factor of 
about $\fgal=10^{-3}-10^3$ above and below this scale.Below we will take both $f_{vir}$ and $f_{gal}$ to be of order one, as their specific value is not relevant to our discussion.

We discussed above that for non-linear structure to form we need $\rho_{\lambda} \ll \rho_{m}$ since vacuum energy can only dominate after the first structures have formed at $t_{vir} < t_{\lambda}$, and obtained the first constraint on $\rho_{\lambda}$ from Eq. \ref{conzero}, which we use below.

Galaxy formation is not well understood and the main mechanisms that contribute to its formation are cooling, gravitational collapse and hierarchical fragmentation \cite{carr,silk,reesost}.Cooling for hydrogen and helium gas clouds can proceed via Brehmsstrahlung or Compton processes, depending on the temperature and density of the collapsing cloud. This is one of the reasons why typical galactic masses and radius sizes are hard to estimate. For example in cooling by Brehmsstrahlung processes galactic baryons cool down to a temperature $T_{cool} = \alpha^{2}m_{e}$ and the mass of the galaxy forming from th egas cloud through this process is $M_{gal} = f^{2}_{gal}M_{*}\alpha_{G}^{-1/2}\alpha^{5} (\frac{m_{p}}{m_{e}})^{1/2}$ \cite{silk}. Then the typical radius of a galaxy with the cooling temperature and mass can be obtained from $T_{cool}, M_{gal}$ from $R_{gal} = \frac{GM_{gal}m_{p}}{T_{cool}}$ and similarly for Compton cooling \cite{silk}. The expression we use here $R_{gal} \simeq R_{max}$ for galaxy sizes, fits the main sequence and it differs by a power of $\alpha_G$ from the Brehmsstrahlung estimates above. The difference may be due to the fragmentation not being included \cite{silk,carr}. In any case, we  use the following expression

\be
R_{gal} = \alpha^{3}(\frac{m_{p}}{m_{e}})^{3/2} \alpha_{G}^{-3/2} 
\label{galaxysize}
\ee

as the typical galaxy size.

The expression of Eqn. \ref{galaxysize} can also be obtained from a crude estimate, using the relation $ (4 \pi/3) R_{gal}^{3} = \frac{M_{gal}}{\rho_{vir}}$, and the expression for $\rho_{vir}/\rho_{eq} \simeq (t_{eq}/t_{vir}^)2$, $a\simeq t^{2/3}$, with $ t_{vir}$ given by Eqns. \ref{virialtime, virialenergy}. Further expressing $t_{eq}, Q$ in terms of the fundamental constants \cite{silk} we have $R_{gal}$ expressed in terms of these constants.

Since overdensities begin to grow at $t_{eq}$ to form a galaxy at later times $\tvir = \timeeq Q^{-3/2} \fvir \,$, from {\it condition 2} we require that the virialized structure can not be larger than the Hubble radius at equality $R_{max} < H^{-1}_{eq} $.   In other words, the radius of the gas cloud which later forms a 
galaxy must be smaller than the horizon size $H_{eq}^{-1}$ at equality time given by Eqn. \ref{equalityhubble} since the horizon mass at equality $M_{eq}$ of Eq. \ref{equalitymass} represents the largest mass scale that can collapse. However, this length scale expands along with a universe that contains dark energy, until the growing
density profile which will become the galaxy decouples from the Hubble flow. Since
the density profile and $Q$ grow as the scale factor $\delta \propto
 a$ the maximum size is rescaled by the factor $1/Q$. Putting this all together, we have this contraint on the size of the galaxy 
 
 \be R_{gal} < \frac{t_{eq}}{Q}.
\label{galacticsize} 
 \ee

Since $Q^{3} \gg \frac{\rho_{\lambda}}{\rho_{eq}}$ from Eq. \ref{conzero},  from Eq. \ref{galacticsize} we have the following

\be
R_{gal} < \frac{t_{eq}}{Q} < t_{eq} \left(\frac{\rho_{eq}}{\rho_{\lambda}} \right)^{1/3}
\label{secondvacbound}
\ee

From Eq. \ref{equalitytime} we know that $t_{eq} = [(\frac{8\pi G}{3}) \rho_{eq} ]^{-1/2}$ and $\rho_{eq} = 2a_{R}T^{4}_{eq}$. Replacing this expression in Eq. \ref{secondvacbound} we finally obtain our important result: a relation that connects the hierarchy of vacuum energy relative to Planck scales, with the hierarchy of forces in the universe given by the ratio of two constants of nature, $\alpha, \alpha_G$. We find 

\be \frac{\rho_{\Lambda}}{M_{pl}^{4}} < \frac{1}{32 \eta^{2}}a_{R}^{-1/2}(\frac{3}{\pi})^{3/2} \left(\frac{\alpha_{G}}{\alpha}\right)^{9/2}(\frac{m_{p}}{m_{e}})^{-9/2} \alpha^{-9/2} (\frac{\Omega_{b}}{\Omega_{m}})^{2} 
\label{finalvacbound}
\ee

 For any habitable universe, the two constants of nature which are the coupling constants for the gravitational ($\alpha_G$) and electromagnetic interactions ($\alpha$), can vary by 6 to 20 orders of magnitude \cite{fred} but for as long as they obey a hierarchy condition given by Eq. \ref{hierarchy}, shown in \cite{fred}. From Eq. \ref{finalvacbound} we have just found that nature selects habitable universes that come with two built in hierarchies, an hierarchy of forces $\frac{\alpha_G}{\alpha} \ll 10^{-34}$ by 34 orders of magnitude or more, and an hierarchy of vacuum energies relative to the Planck energy $\frac{\rho_{\lambda}}{M_{p}} \ll \frac{\alpha_G}{\alpha}^{9/2} \simeq 10^{-120}$  by 120 orders of magnitude or more. Furthermore the extremely suppressed energy of the vacuum relative to Planck energies given by Eq. \ref{secondvacbound} seems to be related to the hierarchy between electroweak and gravitational forces $\frac{\alpha_{G}}{\alpha} \ll 10^{-34}$.

\it{Conclusion and Discussion}\rm

Previously it was shown that under conservative assumptions on structure formation, habitable universes are only those for which there is an hierarchy of forces \cite{fred} given by the ratio of two constants of nature $\frac{\alpha_G}{\alpha} \ll 10^{-34}$. We revisited this problem for the case when the universe is allowed to have a nonzero vacuum energy besides the matter and radiation energy components. We impose a further reasonable constraint on structure formation in universes with dark energy, namely that the size of the gas cloud which later forms a galaxy, be smaller than the horizon size of the universe at equality time. The combined effects of the galaxy size requirement Eqn. \ref{galaxysize}relative to the horizon size at $t_{eq}$ from  the epoch of matter radiation equality until the epoch of vacuum energy domination, in combination with the condition on the hierarchy of forces of Eqn. \ref{hierarchy}, leads to an unexpected condition  on the allowed vacuum energy in habitable universe, making its suppression a function of the hierarchy between the gravitational and electromagnetic forces.

We find it intriguing that the hierarchy of vacuum energy relative to Planck energies should be directly constrained by bounds on the constants of nature and their hierarchy of forces given by Eqn.\ref{finalvacbound}. We showed that the vacuum energy must always be smaller than $10^{-120} M_{planck}^4$ for all habitable universes with $Q < 1$ and for any allowed combination of the the gravitational and fine structure constant $\alpha, \alpha_G$, each of which can vary and draw values in an interval spanning 10 orders of magnitude or more \cite{fred}. Our result Eq.\ref{finalvacbound} in combination with Eq.\ref{hierarchy} means habitable universes will always have a large hierarchy between the vacuum energy and Planck energy. Within that hierarchy, the intriguing relation of dark energy to the constants of nature, still allows it to vary by 30 orders of magnitude from the value of dark energy observed in our universe,  since the electromagnetic and gravitational constants of nature constraining $\rho_{\lambda}$ via Eqn.\ref{finalvacbound}, can independently vary by about 6 to 20 orders of magnitude away from the observed value and still allow universes to be populated by long lived structures \cite{fred}.

For all purposes, a purely constant energy of the vacuum is one of the best contenders we have explaining the acceleration of the universe. If that is the case, then like the Planck mass $M_p$or equivalently Newton's constant $\alpha_G$,  the energy density of the vacuum $\rho_{\lambda}$ adds one more constant of nature and one more hierarchy relative to the gravitational force, in the list of constants of nature and their hierarchies, a universe comes with.
Our finding here is highly suggestive of a connection between the hierarchy of vacuum energy energy relative to Planck energies, and the hierarchy of forces, encapsulated by the ratio of the gravitational and fine structure constants. We are exploring the possibility of an underlying microphysical mechanism that limit contributions to vacuum energy from phase transitions in the early universe, and furthermore which relates the energy of the vacuum to the coupling constants of nature and their hierarchy, in other words to the the Standard Model, in a fundamental way, Eq. \ref{finalvacbound}.  At face value, Eq. \ref{finalvacbound}, provides a relation between coupling constants of the standard model of particle physics and the cosmological constant. We will report our ideas on the possible connection between the in future work. 
\label{sec:conclude}

\acknowledgments

We would like to thank Fred Adams, Evan Grohs, for very useful discussions. 

 LMH acknowledges support from the Bahnson trust fund.




\end{document}